\documentclass[prb, twocolumn, showpacs, superscriptaddress]{revtex4}
\usepackage{amsmath,amsfonts, amsthm}
\usepackage{bm}
\usepackage{mathrsfs}
\usepackage{graphicx, subfigure}
\usepackage{subfigure}
\usepackage{verbatim}
\usepackage{mathrsfs}
\usepackage{color}

\newcommand{\di}{\mathrm{d}}

\renewcommand{\vec}[1]{{\mathbf #1}}

\renewcommand{\vr}{{\vec{r}}}

\renewcommand{\ol}[1]{\overline{#1}}
\newcommand{\comments}[1]{}
\DeclareMathOperator{\sgn}{sgn}

\renewcommand{\emph}[1]{\textit{#1}}

\begin{document}
\title{Criticality in Translation-Invariant Parafermion Chains}
\author{Wei Li}
\affiliation{Physics Department, Arnold Sommerfeld Center for Theoretical Physics,
and Center for NanoScience, Ludwig-Maximilians-Universit\"at, 80333 Munich, Germany}
\author{Shuo Yang}
\affiliation{Max-Planck-Institut f\"ur Quantenoptik, Hans-Kopfermann-Str. 1, D-85748 Garching, Germany}
\affiliation{Perimeter Institute for Theoretical Physics, Waterloo, ON, N2L 2Y5, Canada}
\author{Hong-Hao Tu}
\affiliation{Max-Planck-Institut f\"ur Quantenoptik, Hans-Kopfermann-Str. 1, D-85748 Garching, Germany}
\author{Meng Cheng}
\affiliation{Station Q, Microsoft Research, Santa Barbara, CA 93106-6105, USA}
\date{\today}
\begin{abstract}
	In this work we numerically study critical phases in translation-invariant $\mathbb{Z}_N$ parafermion chains with both nearest- and next-nearest-neighbor hopping terms. 
	The model can be mapped to a $\mathbb{Z}_N$ spin model with nearest-neighbor couplings via a generalized Jordan-Wigner transformation and translational invariance ensures that the spin model is always self-dual. We first study the low-energy spectrum of chains with only nearest-neighbor coupling, which are mapped onto standard self-dual $\mathbb{Z}_N$ clock models. For $3\leq N\leq 6$ we match the numerical results to the known conformal field theory(CFT) identification. We then analyze in detail the phase diagram of a $N=3$ chain with both nearest and next-nearest neighbor hopping and six critical phases with central charges being $4/5$, 1 or 2 are found.  We find continuous phase transitions between $c=1$ and $c=2$ phases, while the phase transition between $c=4/5$ and $c=1$ is conjectured to be of Kosterlitz-Thouless type.
\end{abstract}
\maketitle
\section{Introduction}
In recent years non-Abelian anyons have been a focus of intense theoretical and experimental investigations~\cite{TQCreview}. Their exotic properties have deepened our understanding of quantum many-body phases and also found potential applications in building topological quantum computers. However, quantum phases that host these exotic quasiparticles,  namely non-Abelian topological phases, are usually quite elusive in nature and often require delicate conditions (e.g. complicated forms of many-particle interactions) to occur. Recently, proposals of engineering non-Abelian phases from more conventional materials have greatly stimulated this field of research~\cite{Fu_2008, Zhang2008, Sau_PRL2010, Alicea'12}. For example, Majorana zero modes, being analogues of Ising anyons, have been proposed to exist at ends of semiconductor nanowires in proximity to s-wave superconductors~\cite{1DwiresLutchyn, 1DwiresOreg}, as well as at the magnetic/superconducting domain walls on the edge of two-dimensional topological insulators~\cite{MajoranaQSHedge}. The effort has culminated in the experimental observation of possible signatures of Majorana zero modes in semiconductor/superconductor heterostructures~\cite{Mourik2012, Fink2012, Deng2012, Das2012, Churchill2013}.

Following this line of ideas, it has been proposed that certain extrinsic defects in two-dimensional topologically ordered phases can bind exotic zero modes which are natural generalizations of Majorana zero modes~\cite{Barkeshli_arXiv2013a, Barkeshli_arXiv2013b}. Various physical realizations have been proposed, including magnetic/superconducting domain walls on the edge of two-dimensional fractionalized topological insulators~\cite{Cheng_PRB2012, Clarke_NatCommun2013, Lindner_PRX2012, Klinovaja_arxiv} and dislocations in bilayer quantum Hall systems~\cite{Barkeshli_PRB2013,Barkeshli_PRX2012} or toric-code type models~\cite{You_PRB2012, You_PRB2013, Teo_arxiv2013a}. A common feature among all these seemingly different realizations is that the zero modes can be effectively described by second-quantized operators obeying parafermionic algebra~\cite{Fradkin, Fendley_JStat2012}, which in the simplest case reduce to the well-known  Majorana operators. They are therefore referred as parafermion zero modes subsequently. The parafermion zero modes also exhibit non-Abelian braiding statistics~\cite{Clarke_NatCommun2013, Lindner_PRX2012, Bonderson_PRB2013, Burrello_PRA2013, You_PRB2013, Teo2}, with quantum dimensions squared to an integer.

A recent theoretical development pushes the limit of this engineering approach even further, where it was suggested that even more exotic topological phase, such as the famed Fibonacci phase, can be built by a delicate control of interactions between an array of such defects. The basic fact that underlies this construction is that a chain of interacting $\mathbb{Z}_3$ parafermion zero modes can be tuned to a critical point described by a $\mathbb{Z}_3$ parafermion CFT. Then by assembling many such critical chains together and coupling neighboring chains in an appropriate way~\cite{coupledwire, Neupert_arxiv2014}, a superconducting analogue of the Fibonacci phase can emerge~\cite{Mong_PRX2014, Stoudenmire_unpub, Vaezi_arxiv2013}.

These interesting developments call for a more systematic investigation of the collective behavior of parafermion zero modes, in particular beyond the realm of exact integrability. Unlike Majorana zero modes, a quadratic Hamiltonian of parafermion zero modes is by no means a ``free theory''. They are inherently strongly interacting and even a simple ``quadratic'' Hamiltonian consist of bilinears of parafermion zero modes can exhibit a rich phase diagram. The study of the physics of one-dimensional non-Abelian anyonic chains was pioneered in [\onlinecite{GoldenChains}], where the phase diagram of a chain of interacting Fibonacci anyons was presented, and is subsequently generalized to other anyon models~\cite{GoldenChain2, Gils_PRL2009, Gils_NP2009, Poiblanc_PRB2011, Pfeifer_PRB2012,Gils_PRB2013}. More recently there have been several works on gapped phases of parafermion systems both in one and two dimensions~\cite{Motruk_PRB2013, Bondesan_arxiv2013, Milsted_arxiv, Z3Kitaev}.
In this work we focus on the phase diagram of a translation-invariant quadratic Hamiltonian describing hopping parafermions in one dimension. {By mapping the parafermion Hamiltonian to a $\mathbb{Z}_N$ spin model using a Jordan-Wigner-type transformation~\cite{Fradkin, Fendley_JStat2012}, we see that translation invariance ensures that the spin model is always self-dual, which suggests that these models are critical.} We first analyze the critical phases when there are only nearest-neighbor hoppings. It is well-known that these Hamiltonians can be mapped to self-dual $\mathbb{Z}_N$ clock models~\cite{Fendley_JStat2012} which have been studied thoroughly in the context of classical statistical mechanics.
For $3\leq N\leq 5$ we match the low-energy spectrum with theoretical predictions. In particular, we highlight the subtleties in identifying the CFT spectra due to the non-diagonal CFT partition functions when $N=3$.

We then study the phase diagram of a $\mathbb{Z}_3$ parafermion chain with both nearest-neighbor(NN) and next-nearest-neighbor(NNN) couplings, which leads to a spin model that has not been considered before. Translation invariance still guarantees criticality, but we observe numerically that as the ratio between the NN and NNN couplings are tuned, critical phases with different central charges including $c=4/5, 1, 2$ are realized. We also characterize the phase transitions between these critical phases. We find that the transitions between $c=1$ and $c=2$ phases are continuous.

The paper is organized as follows: in Sec. \ref{sec:model} we introduce the parafermion zero modes and define the model Hamiltonian that is the focus of the paper. We also briefly review the generalized Jordan-Wigner transformation. In Sec. \ref{sec:phases} we establish the phase diagram of $\mathbb{Z}_N$ parafermion chains with NN couplings. In Sec. \ref{sec:z3} we study $\mathbb{Z}_3$ parafermion chains with both NN and NNN couplings. Sec. \ref{sec:conclusion} concludes the paper.

\section{Model and Definition}
\label{sec:model}
Let us first define formally what parafermion zero modes are. A $\mathbb{Z}_N$ parafermion mode is defined as a unitary operator $\gamma$ such that $\gamma^N=1$. For many parafermions, if a certain ordering prescription is chosen, we then have the commutation relation:
\begin{equation}
	\gamma_i\gamma_j=\omega^{\sgn(j-i)}\gamma_j\gamma_i,\: \omega=e^{\frac{2\pi i}{N}}.
  \label{eq:commutation}
\end{equation}
When $N=2$ this is the familiar anti-commutation relation of Majorana zero modes. Notice that $2n$ parafermions can be represented by $N^{n}$-dimensional Hilbert space, therefore in a sense each parafermion zero mode carries ``$\sqrt{N}$''-dimensional states.

\begin{figure}[htpb]
  \begin{center}
	\includegraphics[width=\columnwidth]{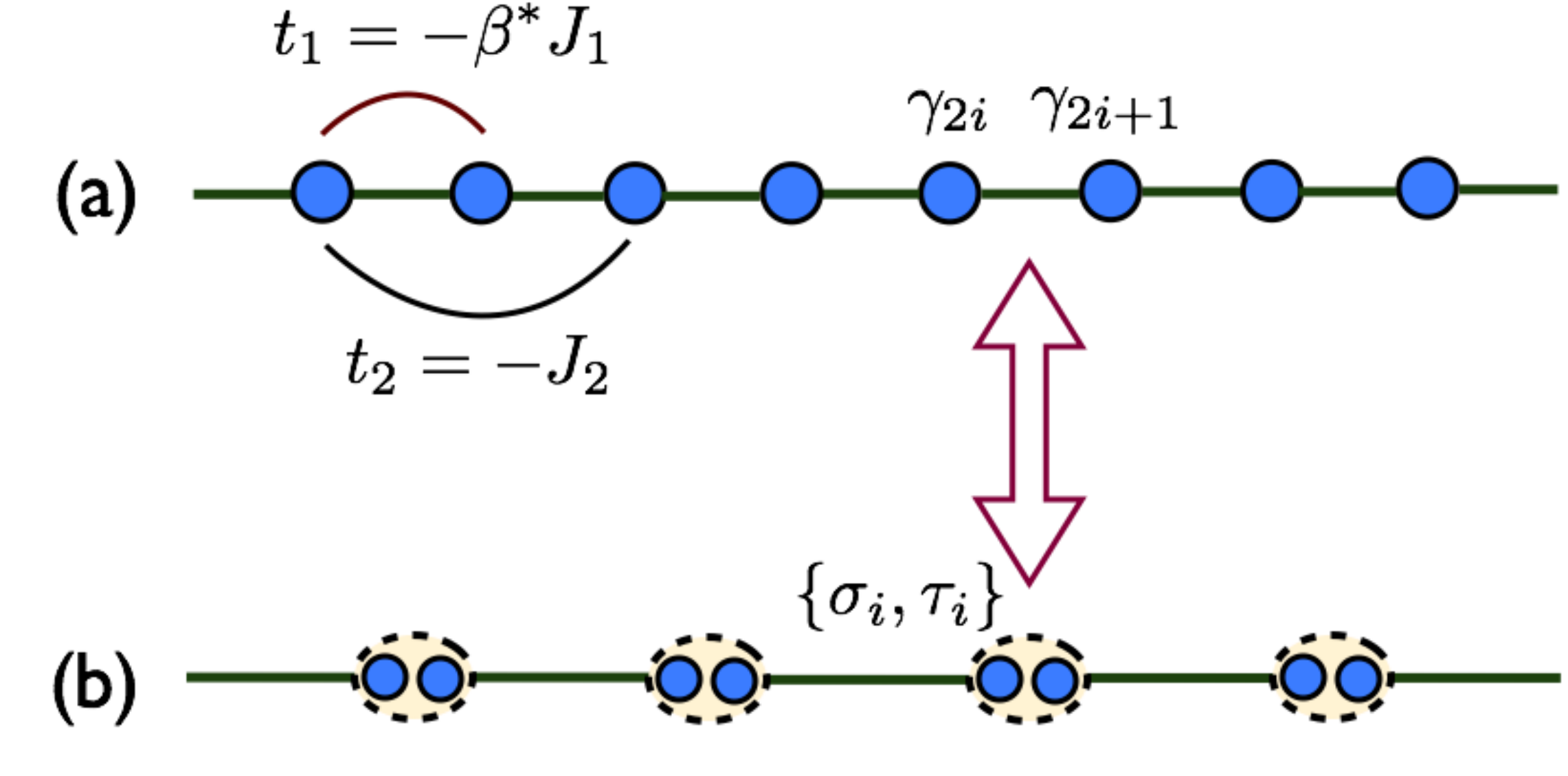}
  \end{center}
  \caption{Illustration of (a) the model of parafermion chain and (b) the corresponding $\mathbb{Z}_N$ spin model.}
  \label{fig:chain}
\end{figure}

In one dimension, lattice sites are naturally ordered. A generic hopping Hamiltonian can be written down:
\begin{equation}
  H=\sum_{ij} (t_{ij}\gamma_i^\dag\gamma_j+\text{h.c.}).
  \label{eq:ham}
\end{equation}
Although the Hamiltonian is quadratic, it is by no means free/non-interacting when $N>2$ due to the parafermionic commutation relation between the operators. If we try to diagonalize the Hamiltonian by Fourier transformation, the commutation relation between the momentum-space modes becomes utterly complicated. Therefore the model is intrinsically a strong-coupling problem. To have a glimpse of the rich physics contained in this model,  we consider hoppings up to the NNN bonds, See Fig. \ref{fig:chain}(a) for an illustration of this parafermion chain.

We notice that the definitions of the operators $\gamma_j$ allow a $\mathbb{Z}_N$ gauge redundancy: $\gamma_j\rightarrow \omega^{n_j}\gamma_j$ where $n_j\in\mathbb{Z}$. The hopping amplitudes also have the same redundancy:
\begin{equation}
  t_{ij}\rightarrow t_{ij}\omega^{n_i-n_j}.
  \label{}
\end{equation}
For example, in an open chain, $t_{ij}$ and $\omega^{|j-i|} t_{ij}$ are identical up to gauge transformations. However, $t_{ij}$ and $-t_{ij}$ in generally are not related for odd $N$. In fact, the ``equivalence classes'' of hopping amplitudes are labeled by gauge-invariant quantities such as $t_{i,i+1}t_{i+1,i+2}t_{i+2,i}=t_1^2t_2^*$. For example, we can perform a gauge transformation $\gamma_i\rightarrow \omega^{-i}\gamma_i$, and $t_1\rightarrow \omega t_1$ and $t_2\rightarrow \omega^2t_2$. Such a transformation can also leave the boundary condition of the parafermions twisted if they sit on a ring. However we mainly focus on open boundary condition and we expect twisted periodic boundary conditions do not affect the major low-energy characterizations of the bulk.

We heavily rely on numerical methods to understand the low-energy physics of this model. In order to carry out numerical simulations, the model is transformed into a $\mathbb{Z}_N$ spin model with a generalized version of Jordan-Wigner transformation~\cite{Fradkin, Fendley_JStat2012}. To be specific, we assume that open boundary condition is imposed. We then define the following tranformation:
\begin{equation}
  \begin{gathered}
	\gamma_{2i}=\sigma_i\prod_{j<i}\tau_j,
	\gamma_{2i+1}=\beta\sigma_i\tau_i\prod_{j<i}\tau_j.
  \end{gathered}
    \label{}
\end{equation}
Here $\sigma_i, \tau_i$ are spin operators that act on a $N$-dimensional Hilbert space for each site, satisfying
\begin{equation}
	\sigma_j^N=\tau_j^N=1, \sigma_j^\dag\sigma_j=\tau_j^\dag\tau_j=1, \sigma_j\tau_j=\omega\tau_j\sigma_j.
  \label{}
\end{equation}
The spin operators on different sites commute.
The constant $\beta$ must satisfy $\beta^N=\omega^{\frac{N(N-1)}{2}}=(-1)^{N-1}$ so that $\gamma_{2i+1}^{N}=1$.

We now apply this transformation to the parafermion chain Hamiltonian with NN and NNN couplings:
\begin{equation}
	H=\sum_i\big(t_1\gamma_i^\dag\gamma_{i+1}+t_2\gamma_{i}^\dag\gamma_{i+2}+\text{h.c.}\big).
	\label{}
\end{equation}

The parafermion bilinears become
\begin{equation}
  \begin{gathered}
	  \gamma_{2j}^\dag\gamma_{2j+1}=\beta\tau_j, \gamma_{2j+1}^\dag\gamma_{2j+2}=\beta^*\omega^*\sigma^\dag_{j}\sigma_{j+1}\\
	  \gamma_{2j}^\dag\gamma_{2j+2}=\sigma^\dag_j\tau_j\sigma_{j+1},\gamma_{2j+1}^\dag\gamma_{2j+3}=\omega^*\sigma_{j}^\dag\sigma_{j+1}\tau_{j+1}
\end{gathered}
  \label{}
\end{equation}
As a result, we obtain the following $\mathbb{Z}_N$ spin model [see Fig. \ref{fig:chain}(b)]:
\begin{equation}
  \begin{split}
	  H=&\sum_{j}(t_1\beta^*\omega^*\sigma_j^\dag\sigma_{j+1}+t_1\beta\tau_j+\text{h.c.})\\
	&-\sum_j(t_2\omega^*\sigma_j^\dag\sigma_{j+1}\tau_{j+1}+t_2\sigma_j^\dag\tau_j\sigma_{j+1}+\text{h.c.}).
  \end{split}
\end{equation}
It is convenient to set $\beta=\omega^{-\frac{N+1}{2}}$ and redefine $t_1=-\beta^* J_1, t_2=-J_2$,
\begin{equation}
  \begin{split}
	  H=&-\sum_{j}(J_1\sigma_j^\dag\sigma_{j+1}+J_1\tau_j+\text{h.c.})\\
	&-\sum_j(J_2\omega^*\sigma_j^\dag\sigma_{j+1}\tau_{j+1}+J_2\sigma_j^\dag\tau_j\sigma_{j+1}+\text{h.c.}).
  \end{split}
    \label{eq-ZN-spin-model}
\end{equation}
We will be working with this form of the Hamiltonian in the rest of the paper and mainly focus on the case where both $J_1$ and $J_2$ are real for simplicity.

Let us examine the symmetries of the Hamiltonian. The parafermion model, as well as the spin model obtained by applying the Jordan-Wigner transformation, both have a global $\mathbb{Z}_N$ symmetry generated by
\begin{equation}
  Q=\prod_{j}\tau_j=\prod_{j}\gamma_{2j}^\dag\gamma_{2j+1}.
  \label{}
\end{equation}
$Q$ can be regarded as the global $\mathbb{Z}_N$ charge. Later the $\mathbb{Z}_N$ quantum numbers will be exploited in the numerical simulation.

We now turn to space-time symmetry. We can define space inversion and time reversal transformations as follows:
\begin{equation}
	\begin{gathered}
	\mathcal{I}:\sigma_j\leftrightarrow \sigma_{-j}, \tau_j\leftrightarrow\tau_{-j}\\
	\mathcal{T}:\sigma_j\leftrightarrow \sigma_j^\dag, \tau_j \leftrightarrow \tau_j
	\end{gathered}
	\label{}
\end{equation}
The $J_2$ term breaks both the inversion and the time-reversal symmetry.

We now show that the spin model obtained in this way is always self-dual. Let us define the dual disorder variables:
\begin{equation}
	\mu_{j}=\prod_{l\leq j}\tau_l, \nu_{j}=\sigma_{j}^\dag\sigma_{j+1}.
  \label{}
\end{equation}
Under the duality transformation, we have
\begin{equation}
	\begin{gathered}
	\sigma_j^\dag\sigma_j\rightarrow \nu_j,
	\tau_j\rightarrow \mu_{j-1}^\dag \mu_{j}\\
	\sigma_j^\dag\sigma_{j+1}\tau_{j+1}\rightarrow \nu_j\mu_j^\dag\mu_{j+1}.
	\end{gathered}
	\label{}
\end{equation}
So the Hamiltonian is invariant.

We notice that the self-duality of the spin model corresponds exactly to the translation invariance of the parafermion model~\cite{Mong_Potts}. Therefore a translation-invariant parafermion chain always maps to a self-dual spin model. Although there is no rigorous proof that self-duality implies criticality, we are not aware of any counterexamples in one-dimensional systems. We will see in the following sections that our model indeed exhibits criticality.

\section{$\mathbb{Z}_{N}$ model with only NN couplings}
\label{sec:phases}

In this section, we start from the $\mathbb{Z}_{N}$ model (\ref {eq-ZN-spin-model}) with $J_{2}=0$, where only NN couplings are present. For $J_{1}=1$, the $\mathbb{Z}_{N}$ parafermion chain maps exactly to the self-dual $\mathbb{Z}_{N}$ clock model. The study of the phase diagram of $\mathbb{Z}_{N}$ clock model has a long history. Utilizing a field theoretical approach, it has been argued that \cite{lecheminant2002} the quantum model in one dimension can be related to the classical planar XY model with $\mathbb{Z}_{N}$ symmetry-breaking fields in two dimensions in certain anisotropic limit. The Euclidean action of the classical model is equivalent to that of a sine-Gordon model given by~\cite{lecheminant2002}
\begin{equation}
\mathcal{S}=\frac{1}{2}\int \mathrm{d}^{2}\vr\,[(\nabla \varphi)^2+g\cos
(\sqrt{N}\varphi)+\tilde{g}\cos (\sqrt{N}\theta)],  \label{eq:Sine-Gordon}
\end{equation}%
where $\varphi (\vr)$ is a bosonic field and $\theta (\vr)$ is the
conjugate field. The two cosine potentials in (\ref{eq:Sine-Gordon}) always have
the same scaling dimensions, thus competing with each other. The duality transformation of the spin model corresponds to $\varphi\leftrightarrow \theta$ in the field theory. Hence when $g=\tilde{g}$ the field theory is self-dual. It is useful to first consider the weak-coupling limit and perform a renormalization group analysis of the perturbations to the Gaussian fixed point. For $N<4$, the two perturbations $\cos \sqrt{N}\varphi$ and $\cos\sqrt{N}\theta$
are both relevant. So they drive the theory to a new strong-coupling fixed point whose central charge is less than $1$ according to Zamolodchikov's $c$-theorem~\cite{ctheorem}. Therefore the infra-red fixed-point is necessarily a CFT minimal model.
It is known that the new fixed-point is described by an Ising CFT with central charge $c=1/2$ for $N=2$,
and a $\mathbb{Z}_{3}$ parafermion CFT with central charge $c=4/5$ for $N=3$.
For $N=4$, the perturbation is marginal and the low-energy fixed-point will be identified below. For $N\geq 5$ the cosine perturbations
in (\ref{eq:Sine-Gordon}) become irrelevant, thus the low-energy fixed point is again Gaussian.

We now come back to lattice models. For $N=3$, the clock model (\ref{eq-ZN-spin-model}) coincides with the famous $3$-state Potts model, which has the special property
of being integrable \cite{Sierra_book}. $J_1=1$ will be called the ferromagnetic(FM) coupling, since the $\mathbb{Z}_3$ spins are aligned in the same direction in the ground state of the Hamiltonian with just the term $-J_1\sum_j (\sigma_j^\dag\sigma_{j+1}+\sigma_{j+1}^\dag\sigma_j)$, and  correspondingly $J_1=-1$ the antiferromagnetic(AF) coupling. For the FM case, one can deduce from the Bethe ansatz solutions
that the chain has gapless excitations \cite{albertini1992} and the low-energy effective theory is the $\mathbb{Z}_{3}$
parafermion CFT \cite{dotsenko1984}, also
known as the minimal model $\mathcal{M}(6,5)$ with a central charge $c=4/5$.
 However, the field content of the three-state Potts criticality differs from the genuine $\mathcal{M}(6,5)$ CFT. Only $6$ primary fields out of the $10$ with conformal
dimensions $h=0,2/5,7/5,3,1/15,2/3$ are responsible for the ferromagnetic
three-state Potts model. In fact, the field content is completely specified by the following non-diagonal modular-invariant partition function~\cite{cardy1986}:
\begin{equation}
Z_{F} =|\chi _{0}+\chi_{3}|^2
+|\chi _{\frac{2}{5}}+\chi_{\frac{7}{5}}|^2
+2|\chi _{\frac{1}{15}}|^2+2|\chi _{\frac{2}{3}}|^2,
\label{eq:Z3character}
\end{equation}%
where $\chi_{h}=\mathrm{tr}_{h}(q^{L_{0}-c/24})$ is the holomorphic CFT character with $\mathrm{tr}_h$ the trace in the conformal block $h$ (labeled by the conformal dimension).

For the $\mathbb{Z}_{3}$ clock model with antiferromagnetic coupling, it was proposed in [\onlinecite{McCoyZ4}] based on Bethe ansatz that the critical theory
should be the $\mathbb{Z}_{4}$ parafermion CFT with central
charge $c=1$. The
partition function relevant for the antiferromagnetic three-state Potts
model has been shown to be the non-diagonal combination of characters
\cite{gepner1987}%
\begin{equation}
Z_{A} =|\chi _{0}+\chi _{1}|^2
+4|\chi _{\frac{3}{4}}|^2
+2|\chi _{\frac{1}{3}}|^2+2|\chi _{\frac{1}{12}}|^2,
\label{eq:Z4character}
\end{equation}%
where five primary fields with conformal dimensions $h=0,3/4,1,1/3,1/12$
show up. Notice that if we represent the $\mathbb{Z}_4$ parafermion CFT as the coset $\mathbb{SU}(2)_4/\mathbb{U}(1)$, then only fields with integer $\mathbb{SU}(2)$ spins appear in the partition function. This implies that the CFT can be obtained from $\mathbb{SU}(2)_4/\mathbb{U}(1)$ by ``condensing'' the highest spin primary fields in $\mathbb{SU}(2)_4$, which results in $\mathbb{SU}(3)_1$ theory~\cite{BaisPRB2009}.  With this perspective, the CFT of the antiferromagnetic $\mathbb{Z}_3$ Potts chain should better be described as $\mathbb{SU}(3)_1\times\overline{\mathbb{U}(1)}_2\simeq \mathbb{U}(1)_6$~\footnote{We thank E. Ardonne for very helpful correspondence on the CFT of antiferromagnetic $\mathbb{Z}_3$ Potts model. }. 

\begin{figure}[htpb]
  \begin{center}
	\includegraphics[width=0.95\columnwidth]{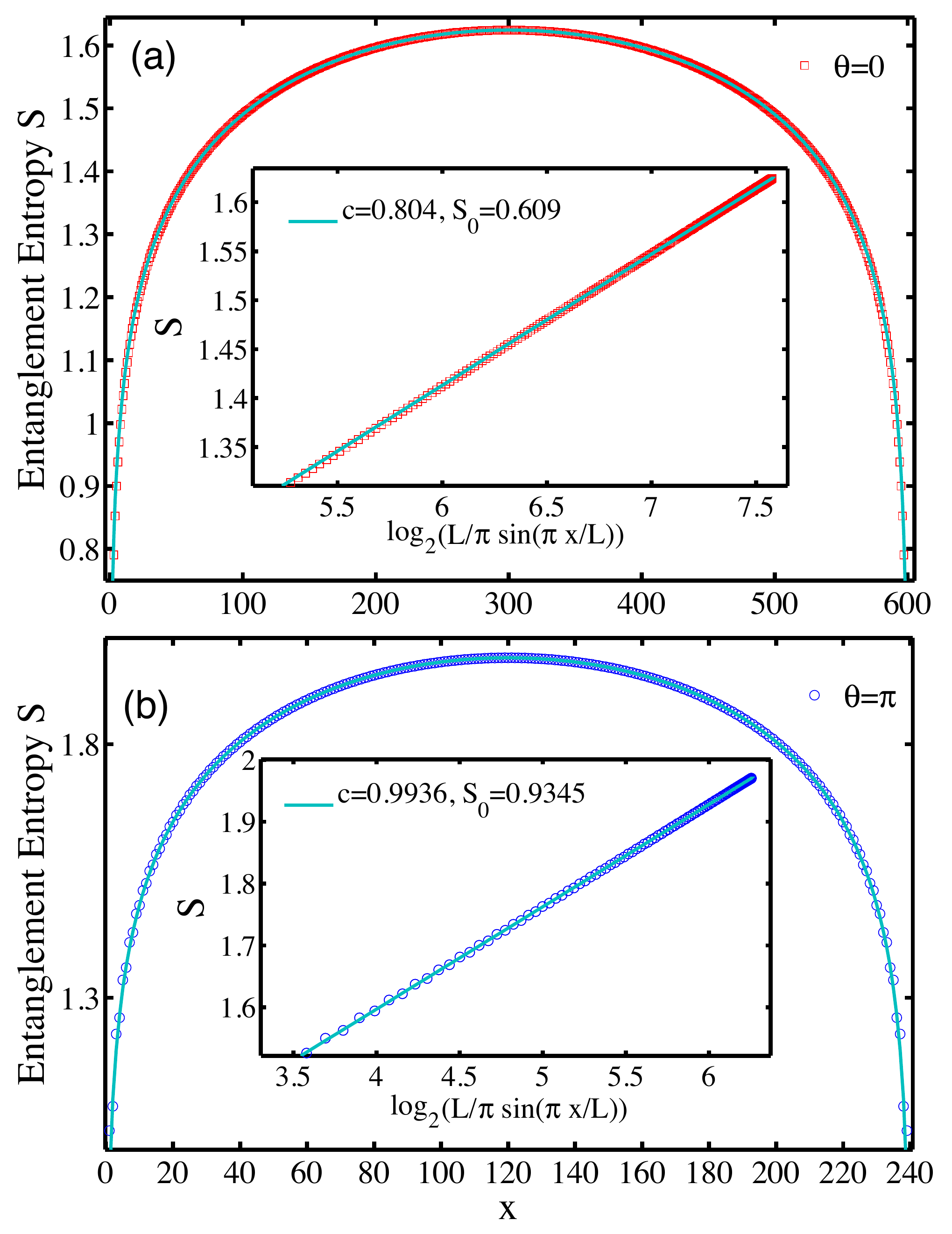}
  \end{center}
  \caption{Block entanglement entropy of the $\mathbb{Z}_3$ FM($\theta=0$) and AF($\theta=\pi$) Potts model. (a) $\theta=0$ with $c=\frac{4}{5}$, and (b) $\theta=\pi$ with $c=1$. Open boundary conditions are adopted in both cases.}
  \label{fig:entropy}
\end{figure}

In order to verify these field theoretical predictions, we perform numerical
simulations for $\mathbb{Z}_{N}$ clock models (\ref{eq-ZN-spin-model})\
with only nearest-neighbor interactions, based on density-matrix renormalization
group (DMRG) and exact diagonalization (ED) techniques. The numerical
methods that we adopt here also form the basis for our further
investigations of more complicated models in subsequent sections. In the
DMRG method, we approximate the ground states and sometimes also several
low-lying excited states of the Hamiltonian with open boundary conditions by
matrix-product states. For the ground states of 1D critical quantum chains
with length $L$ and open boundaries, it has been shown\cite{Holzhey,Vidal,Calabrese-Cardy} that the von Neumann entanglement entropy of
a block of $x$ consecutive spins scales as
\begin{equation}
S=\frac{c}{6}\log_2 {\left( \frac{L}{\pi }\sin \frac{\pi x}{L}\right) }+S_{0},
\label{eq:Cardy}
\end{equation}%
where $c$ is the central charge of the CFT and $S_{0}$ is a non-universal
constant. In the case of periodic boundary condition, the coefficient of the logarithmic scaling of entanglement entropy in (\ref{eq:Cardy}) should be modified to $c/3$. In DMRG calculations, we fit the numerically computed von Neumann
entropy with this formula, from which one can read off the central charge $c$%
.

Once $c$ is determined, we can compute the energy spectra of a finite-size chain with periodic boundary condition and further constrain the conformal dimensions of primary fields in the CFT. This is based on the following result:
  For a 1D critical chain described by a CFT,
the energy spectra are given by \cite{affleck1986a,blote1986}
\begin{equation}
E=\varepsilon _{\infty }L-\frac{\pi vc}{6L}+\frac{2\pi v}{L}(h+\ol{h}%
+n+\ol{n}),  \label{eq:finite-size}
\end{equation}%
where $\varepsilon _{\infty }$ is the ground-state energy per site in the
thermodynamic limit, $v$ is the sound velocity, $h$ and $\ol{h}$ are
conformal dimensions of the CFT primary fields, and $n$ and $\ol{n}$ are
non-negative integers. In practice, we can find $v$ accurately from the finite-size scaling of the ground state energy. Then comparing the numerically
computed energy spectra with (\ref{eq:finite-size}) allows to extract conformal dimensions of the CFT primary fields,
which are characteristic quantities for identifying the CFT.

Notice that caution should be taken when one tries to extract the holomorphic conformal dimension $h$ from \eqref{eq:finite-size}. From the excited energy spectra we can only obtain $h+\ol{h}$ directly. If the partition function is diagonal, all CFT states have zero conformal spins meaning $h=\ol{h}$. However, in the present case the relevant partition functions of both $\mathbb{Z}_3$ and $\mathbb{Z}_4$ parafermion CFTs are non-diagonal, which means that operators with non-zero conformal spins appear in the spectrum. For example, for the $\mathbb{Z}_4$ parafermion CFT one should find two degenerate levels corresponding to $(h,\ol{h})=(1,0),(0,1)$, and the other levels should all be diagonal meaning that $h=\ol{h}$ (modulo the $n$ and $\ol{n}$ shifts ).

\begin{figure}[tbph]
\centering
\includegraphics[width=\columnwidth]{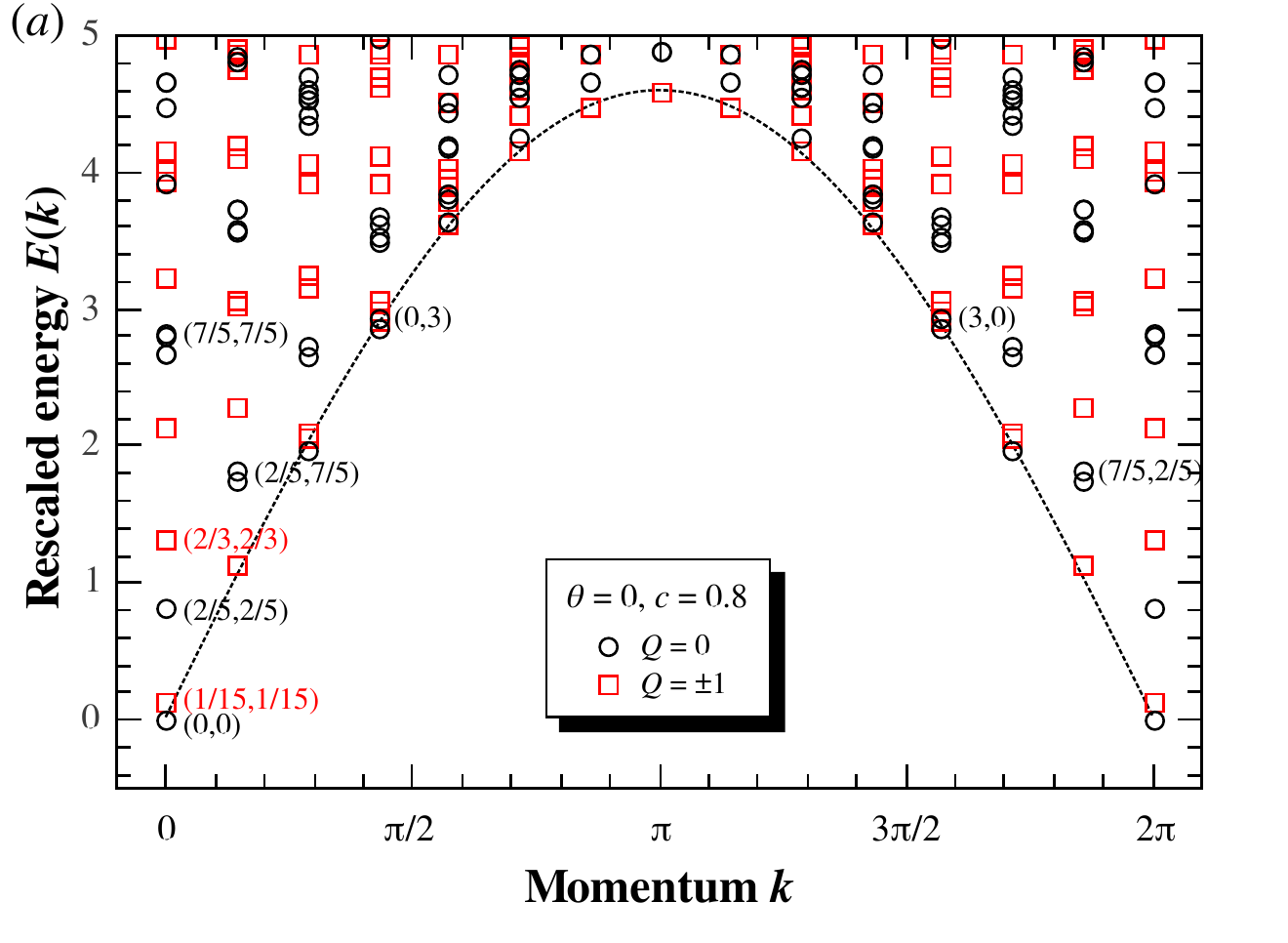} \includegraphics[width=
\columnwidth]{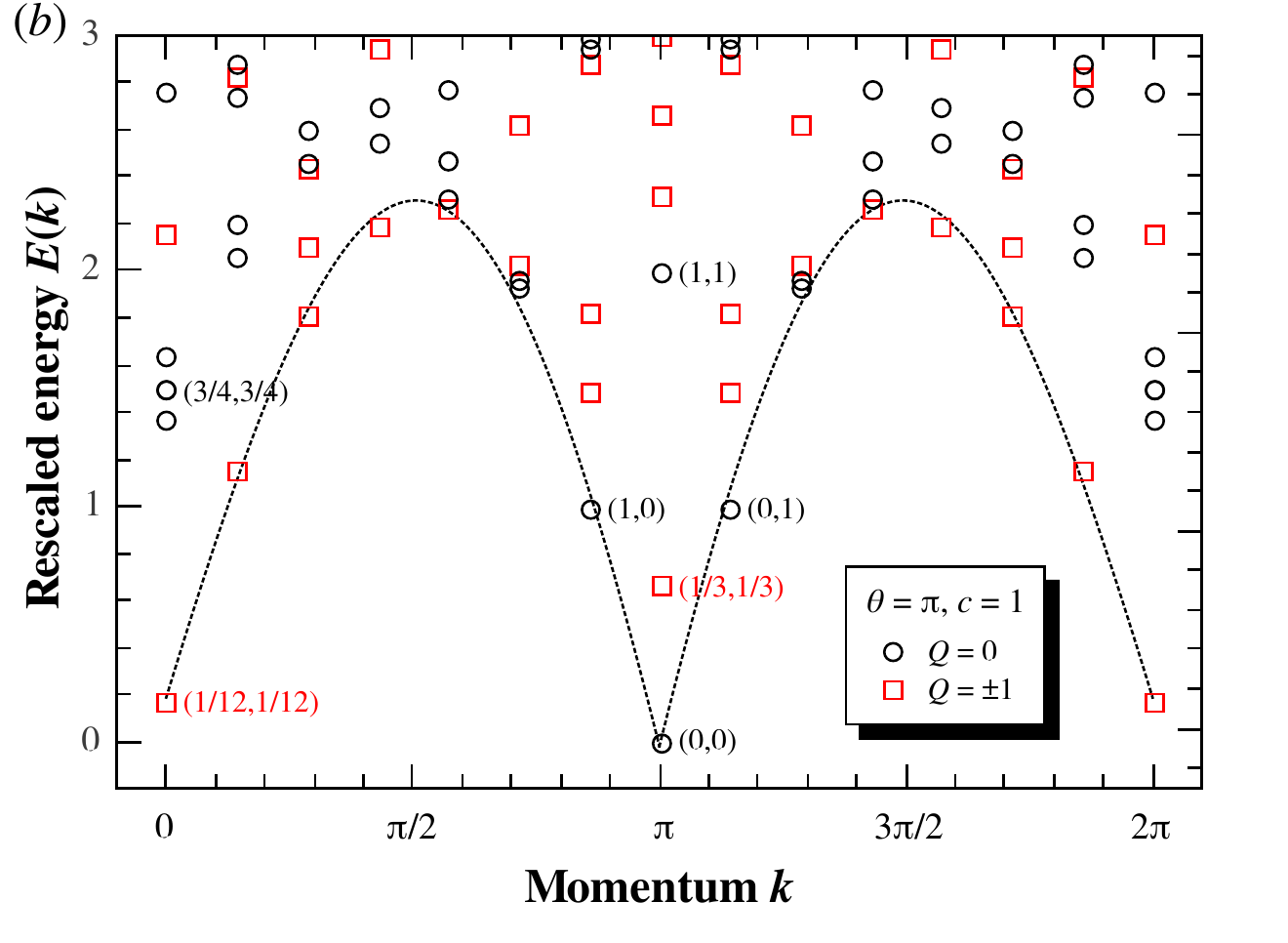}
\caption{Low-energy spectra of the quantum $\mathbb{Z}_{3}$ Potts
model of size $L=14$ with periodic boundary conditions for (a) the ferromagnetic coupling $J_{1}=1$ and (b)
the antiferromagnetic coupling $J_{1}=-1$. The spectra have been shifted and
rescaled by the exact values of the ground-state energy and the sound
velocity according to Eq. (\protect\ref{eq:finite-size}), so that the
comparison to the CFT predictions is more transparent. The open circles
(squares) denote the energy levels with $\mathbb{Z}_{3}$ quantum number $Q=0$
($Q=\pm 1$). The energy levels corresponding to the CFT primary fields are
labeled by their conformal dimensions $(h,\bar{h})$.}
\label{fig:ed}
\end{figure}

For the ferromagnetic $\mathbb{Z}_{3}$ Potts model, we confirm that the central
charge is well fitted to $c=4/5$ in our DMRG calculations. The numerically
computed von Neumann entanglement entropy is shown in Fig. \ref{fig:entropy}
(a). The (rescaled)\ finite-size spectrum of a periodic chain with $L=14$
sites is shown in Fig. \ref{fig:ed}(a). Using Eq. (\ref{eq:finite-size}),
the CFT primary fields appearing in (\ref{eq:Z3character}) are found in
the low-lying spectrum, including the non-diagonal combinations $(3,0)$ and $(0,3)$, see Fig. \ref{fig:ed}(a). Our extrapolated ground-state energy per site $
\varepsilon _{\infty }= -2.43599$ (from DMRG calculations) and sound velocity $v=2.58441$ (from ED results with size $L=14$)
both agree very well with the exact values $\varepsilon _{\infty }=-\frac{2
\sqrt{3}}{\pi }-\frac{4}{3}$ and $v=\frac{3\sqrt{3}}{2}$ from the Bethe
ansatz solution \cite{albertini1992}.

For the antiferromagnetic $\mathbb{Z}_{3}$ Potts model, we find $c=1$ from the
numerical fit of entanglement entropy [see Fig. \ref{fig:entropy}(b)]. We
have observed that the ground-state energy has an even-odd dependence on the
system size, so we extract conformal dimensions from a chain with even
system size $L=14$ [see Fig. \ref{fig:ed}(b)]. This is in agreement with the ED ground state being located at momentum $\pi$. Our numerical results confirm
the $\mathbb{U}(1)_6$ CFT prediction for the antiferromagnetic
three-state Potts model, as well as the relevant primary fields in the
partition function (\ref{eq:Z4character}). Moreover, the numerically computed
ground-state energy per site $\varepsilon _{\infty }=-1.816071$ (from DMRG calculations) and
sound velocity $v=1.2883$ (from ED results with size $L=14$) are also in very good agreement with the
exact values $\varepsilon _{\infty }=-\frac{\sqrt{3}}{\pi }-\frac{3\sqrt{3}}{
2}+\frac{4}{3}$ and $v=\frac{3\sqrt{3}}{4}$ \cite{albertini1992}. All low-lying excited levels can be matched up with CFT predictions, including two non-diagonal combinations $(1,0)$ and $(0,1)$.

\begin{table}
	\centering
	\begin{tabular}{|c|c|c|c|}
		\hline
		$N$ & Coupling & CFT & Remarks\\
		\hline
		$2$ & AF/FM & Ising & \\
		\hline
		$3$ & FM & $\mathbb{Z}_3$ PF &  non-diagonal partition function\\
		\hline
		$3$ & AF & $\mathbb{U}(1)_6$ &  non-diagonal partition function\\
		\hline
		$4$ & AF/FM & $\mathbb{U}(1)_4/\mathbb{Z}_2$ & $R=2$\\
		\hline
		$\geq 5$ & AF/FM & $\mathbb{U}(1)_{2N}$ & $R=\sqrt{2N}$\\
		\hline
	\end{tabular}
	\caption{Summary of the low-energy CFT descriptions of the $\mathbb{Z}_N$ clock model. ``PF'' is short for parafermion CFT. $R$ is the compactification radius of the $\mathbb{U}(1)$ boson CFT. }
	\label{tab:Zn}
\end{table}

Now we turn to $N>3$. We confirm that, for $N=4,5,6$, the central
charges of the $\mathbb{Z}_{N}$ clock models with $J_{1}=\pm 1$ are all equal to $1$, in
consistent with the field theoretical prediction based on \eqref{eq:Sine-Gordon}. In addition, we have extracted the conformal dimensions
of corresponding primary fields from the low-lying excited states. For $N=4$%
, the lowest six primary fields have conformal dimensions $%
h=1/16,1/16,1/8,1/2,1/2,9/16$, in perfect agreement with the $\mathbb{Z}_{2}$
orbifold of a $\mathbb{U}(1)$ boson compactified on a circle of radius $R=2$~\cite{orbifold}, which is
just two copies of Ising CFTs. For $N=5$, regardless of the sign of the coupling the lowest two conformal
dimensions read $h=1/20$ and $1/5$, in agreement with the CFT of a
compactified boson on a circle of radius $R=\sqrt{10}$~\cite{BorisenkoPRE2011}. Thus, we expect that
for $N\geq 5$ all $\mathbb{Z}_{N}$ clock models with $J_{1}=\pm 1$ are described by
$c=1$ free-boson CFT with a compactification
radius $R=\sqrt{2N}$. We summarize these results in Table \ref{tab:Zn}.

\section{$\mathbb{Z}_3$ model with up to NNN couplings}
\label{sec:z3}
\begin{figure}[htpb]
  \begin{center}
	\includegraphics[width=0.8\columnwidth]{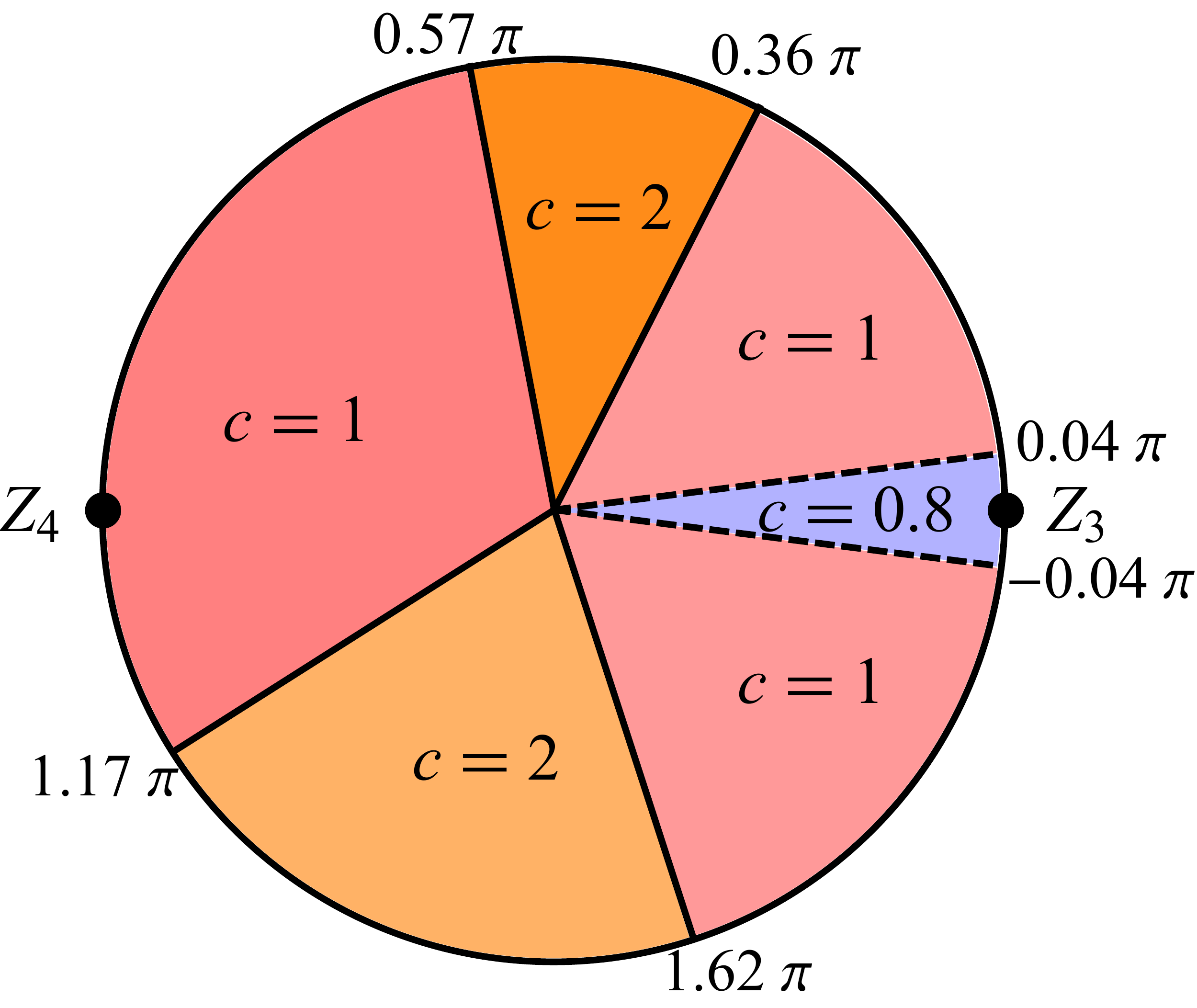}
  \end{center}
  \caption{Phase diagram of the $\mathbb{Z}_3$ Potts model. There are six critical phases labeled by different central charges. There are two exactly solvable points at $\theta=0$ and $\theta=\pi$~\cite{Sierra_book}, which extend to a $c=\frac{4}{5}$ phase and a $c=1$ phase, respectively. In addition, there are two $c=2$ critical phases, roughly centered around the $\theta=\frac{\pi}{2}$ and $\theta=\frac{3\pi}{2}$ points. There also exist two $c=1$ phases between the $c=2$ phases and the $c=\frac{4}{5}$ phase. The positions of transition points separating the $c=1$ and $c=2$ phases are determined by locating the positions of peaks in second-order energy derivatives. The boundaries between $c=\frac{4}{5}$ and $c=1$ phases are more subtle and can be extracted from the entanglement data.}
  \label{fig:pdiag}
\end{figure}

In this part we present the phase diagram of a $\mathbb{Z}_3$ parafermion chain with NN and NNN hoppings, which is summarized in Fig. \ref{fig:pdiag}. We parametrize the two hopping strengths by $J_1=\cos\theta$ and $J_2=\sin\theta$. The Hamiltonian is solved numerically by the DMRG method with open boundary conditions. As expected, the whole phase diagram are filled by critical phases. This is readily seen by calculating the ground state entanglement entropy as a function of the block size $x$ and fitting it with Eq. \eqref{eq:Cardy}. From the scaling we also read off the central charge $c$ of the critical phase.

\begin{figure}[htpb]
  \begin{center}
	\includegraphics[width=0.95\columnwidth]{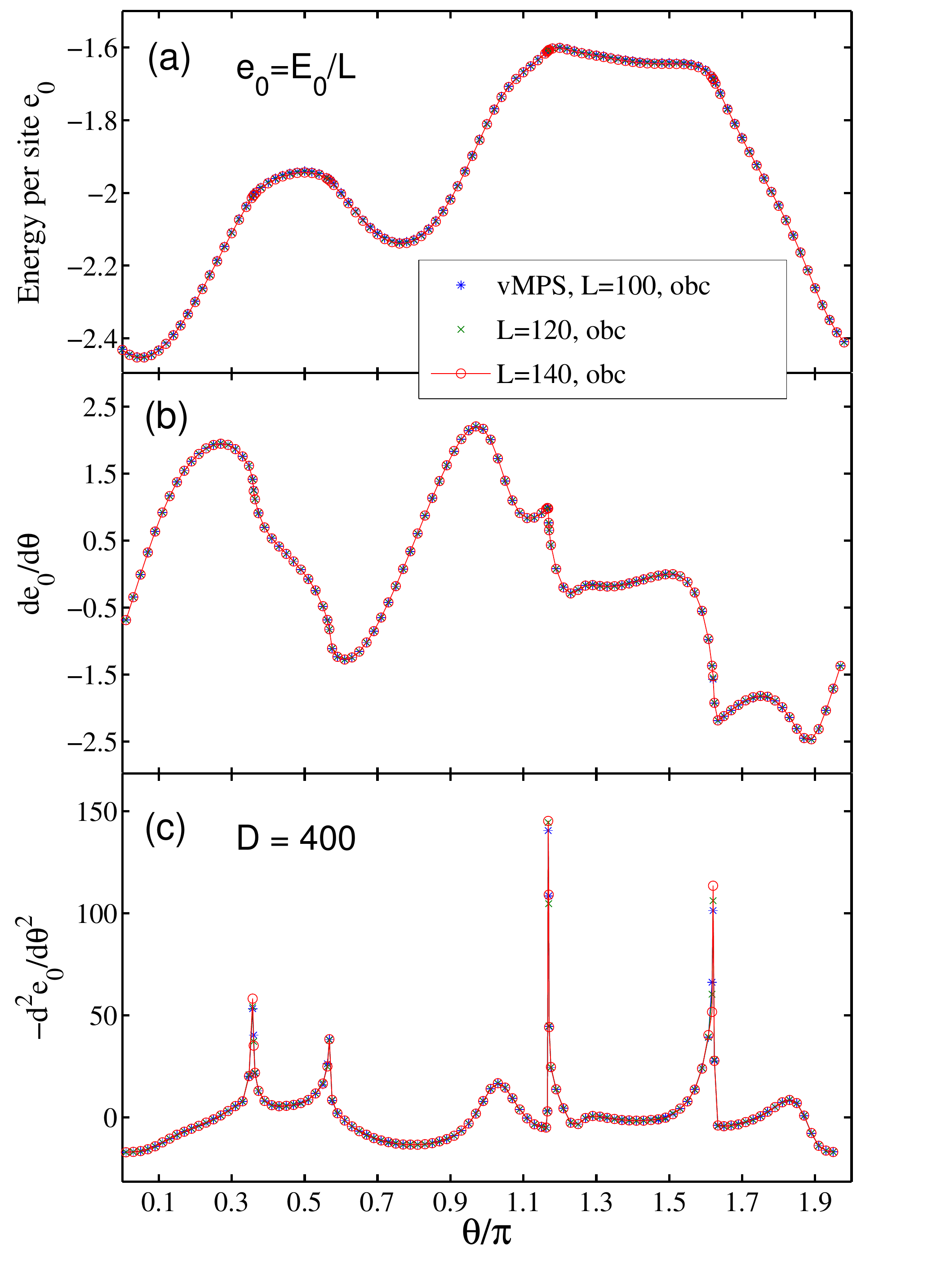}
  \end{center}
  \caption{Ground-state energy per site $e_0$ and its (first- and second-order) derivatives with respect to the parameter $\theta$.  $e_o$ and its derivatives are seen to converge with increasing system sizes nearly everywhere, except for $\theta$ in the vicinity of transition points (diverging peaks of $\di^2e_0/\di \theta^2$). The four peaks (at $0.356\pi$, $0.572\pi$, $1.168 \pi$, and $1.624\pi$) in the $\di^2e_0/\di\theta^2$ clearly signal continuous transitions.}
  \label{fig:energy}
\end{figure}

\begin{figure}[htpb]
  \begin{center}
	\includegraphics[width=0.95\columnwidth]{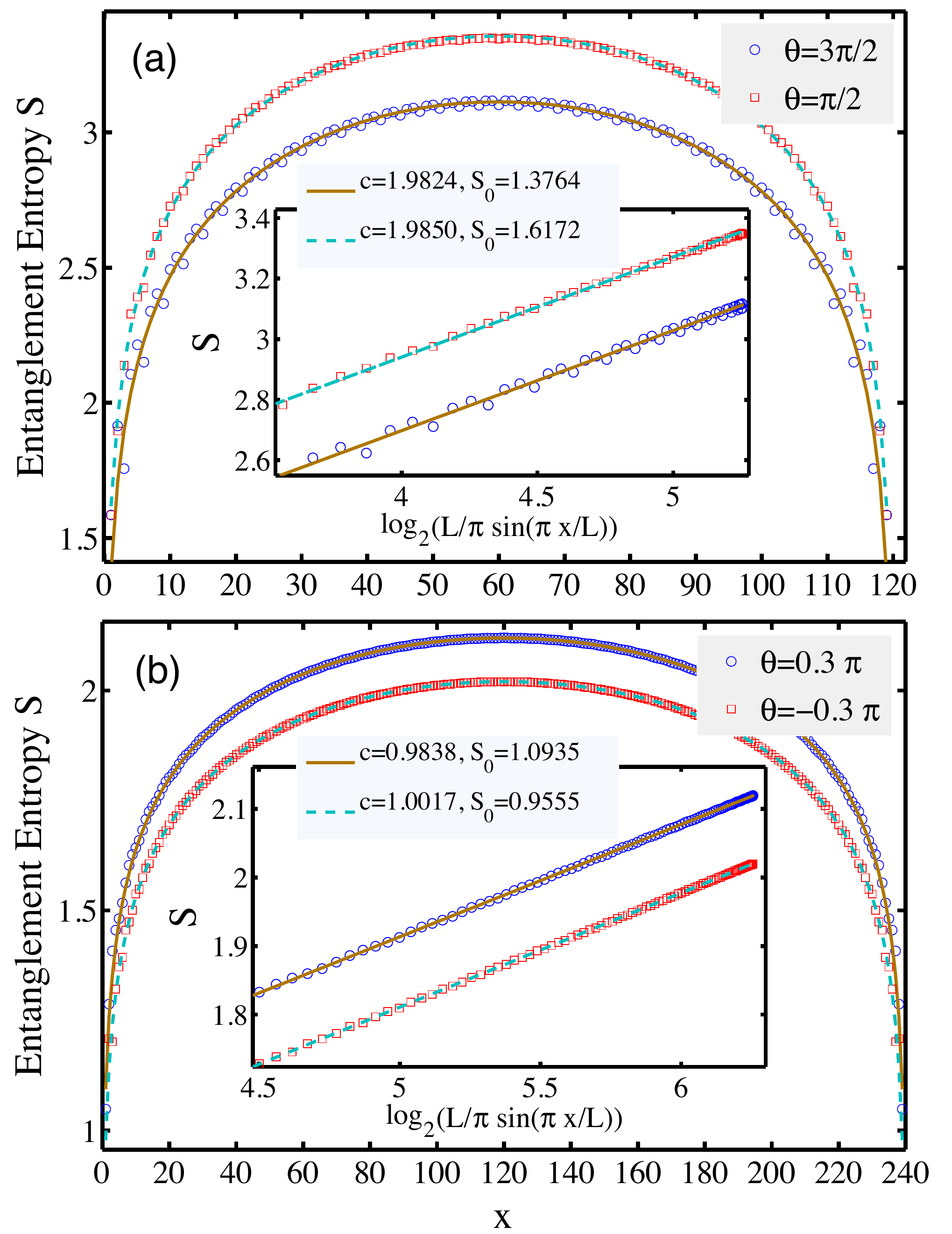}
  \end{center}
  \caption{Block entanglement entropy: (a) $\theta=\pi/2, 3\pi/2$ with $c=2$; (b) selected points in the two phases with $c=1$ above and below the $c=\frac{4}{5}$ phase ($0.05 \pi < \theta < 0.36 \pi$, and $-0.05 \pi < \theta < -0.4\pi$).}
  \label{fig:c=1}
\end{figure}

To accurately pin down the phase boundaries, we first calculate numerically the ground-state energy density (i.e. energy per site) and its first- and second-order derivatives with respect to $\theta$ to locate the phase transition points which at the same time reveal the nature of the phase transitions. In Fig. \ref{fig:energy} we show the energy per site $e_0$, its first and second-order derivatives with respect to $\theta$ as a function of $\theta$. One can clearly see that the first-order derivative $\frac{\di e_0}{\di \theta}$ is continuous and there are discontinuities in the second-order derivative $\frac{\di^2 e_0}{\di \theta^2}$ at $\theta\approx 0.36\pi, 0.57\pi,1.17\pi, 1.62\pi$. We then calculate the central charge in different regions of the phase diagram in order to identify the phases. We show the block entanglement entropy for selected points in Fig. \ref{fig:c=1}. This also provides an alternative check of phase boundaries. We find that the phase transitions between $c=1$ and $c=2$ phases are very likely to be continuous, and in these cases the two ways of obtaining the phase boundaries agree with each other perfectly.

\begin{figure}[htpb]
  \begin{center}
	\includegraphics[width=0.95\columnwidth]{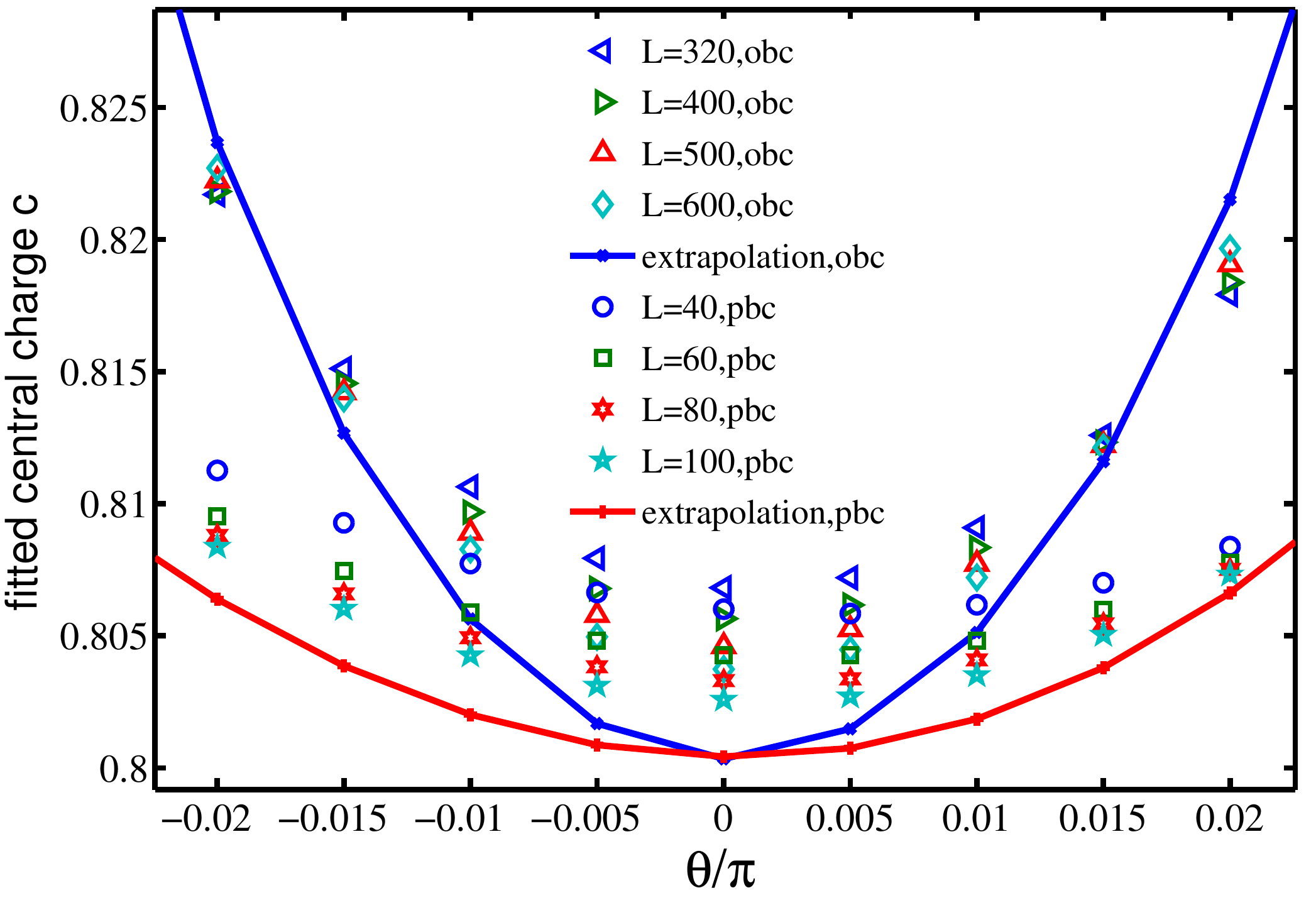}
  \end{center}
  \caption{The fittings of the central charge from the scaling of block entanglement entropies around the $\mathbb{Z}_3$ point $\theta=0$. Both OBC and PBC are used. The results are consistent with the existence of a finite region with $c\approx \frac{4}{5}$, roughly ranging from $\theta=-0.03 \pi$ to $\theta=0.03 \pi$. }
  \label{fig:Z3}
\end{figure}

\begin{figure}[htpb]
  \begin{center}
	\includegraphics[width=0.9\columnwidth]{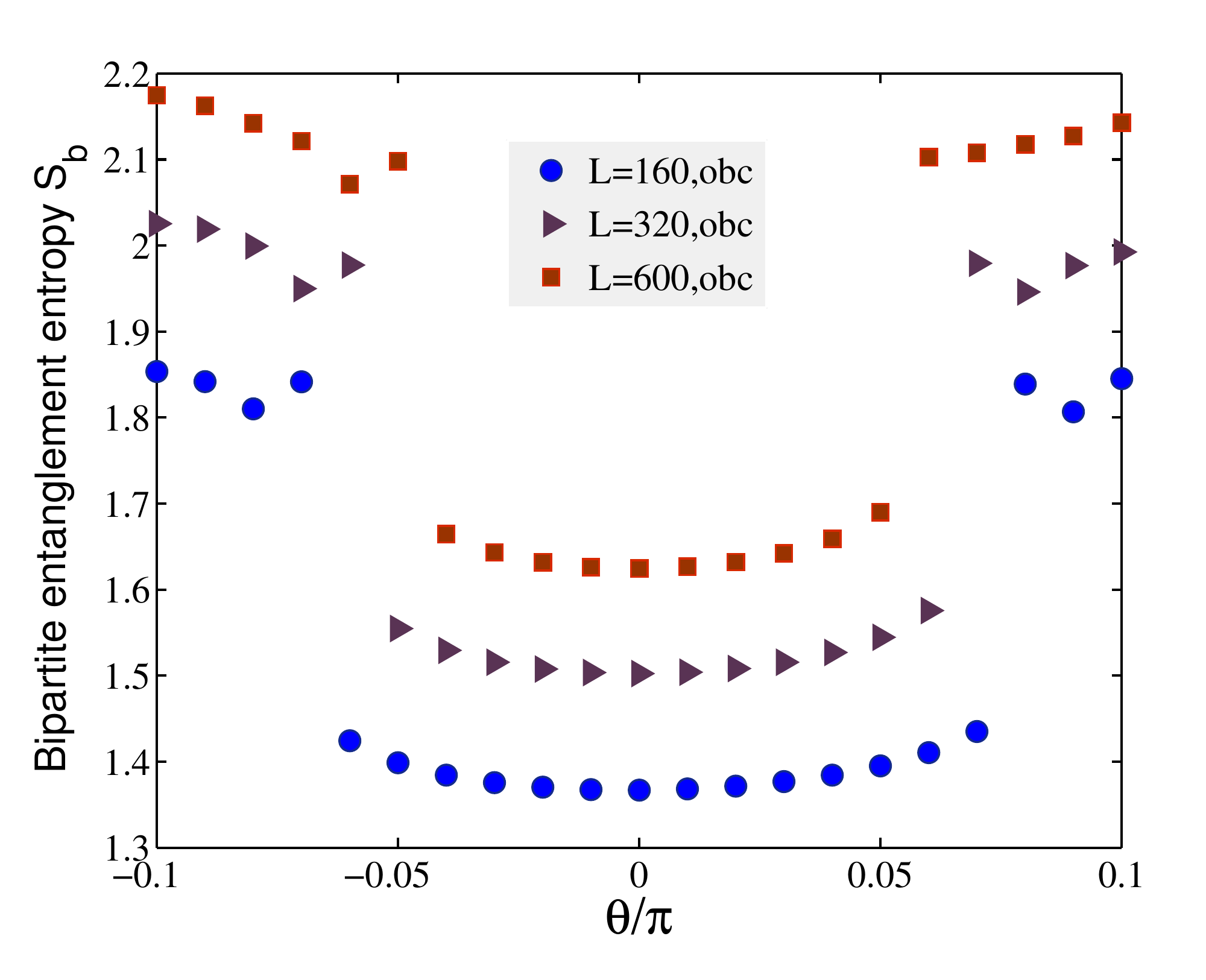}
  \end{center}
  \caption{Bipartite entanglement entropy ($S_b$) by cutting the chain in the middle. $S_b$ shows a clear jump (at around $\theta \approx \pm 0.05 \pi$) between $c=\frac{4}{5}$ and $c=1$ regions. MPS bond dimension is set to $D=300$ in this calculation.}
  \label{fig:bipartite entanglement}
\end{figure}

However, the energy and its derivatives do not show any features near the ``transition'' between the $c=\frac{4}{5}$ phase and the neighboring $c=1$ phases. This part of the phase diagram near $\theta=0$ is particularly relevant to the recent studies of the Fibonacci phase~\cite{Mong_PRX2014,Vaezi_arxiv2013, Vaezi_fib2014}, so we carefully perform finite-size scaling of the central charge to map out the phase diagram in this region, see Fig. \ref{fig:Z3}. We confirm that the $c=\frac{4}{5}$ $\mathbb{Z}_3$ parafermion CFT region extends roughly from $-0.04 \pi$ to $0.04 \pi$, beyond which it is taken over by $c=1$ phases. In order to confirm there is indeed a phase transition, we calculate the bipartite entanglement entropy around the ``transition'' point and observe a clear jump as shown in Fig. \ref{fig:bipartite entanglement}, which implies a dramatic change of the ground state wavefunction between the $c=\frac{4}{5}$ and $c=1$ phases.

Regarding the nature of the transition between the $c=\frac{4}{5}$ and $c=1$ phases, there can be three possibilities theoretically, all consistent with the numerical results: (1) The singularity appears in third- or higher-order derivatives. (2) There is a Kosterlitz-Thouless transition where the derivatives of the energy with respect to the tuning parameter are smooth to all orders. (3) It is a crossover instead of a phase transition. Although we are limited by the accuracy of numerical simulations, we believe a third-order phase transition is not very likely. The third option is also unfavored due to the abrupt change in the ground state entanglement. We therefore conjecture that the phase transition is of the Kosterlitz-Thouless type.

Interestingly, near the $\theta=\frac{\pi}{2},\frac{3\pi}{2}$ points where there are only NNN couplings one may naively think that the chain can be decoupled as two copies of the model with only NN hopping. This expectation is however not true. Due to the unusual commutation algebra (\ref{eq:commutation}) between the parafermions, the two ``copies'' do not commute and are still highly entangled. This is quite different from the case of the Majorana hopping model($N=2$ parafermion chain). When there are only NNN coupling the even sites and the odd sites decouple from each other and form two $c=\frac{1}{2}$ CFTs. In the present case, for both AF and FM NNN couplings, we find $c=2$, obtained by the entanglement fitting in Fig. \ref{fig:c=1}(a).

\begin{figure}[htpb]
  \begin{center}
	\includegraphics[width=0.9\columnwidth]{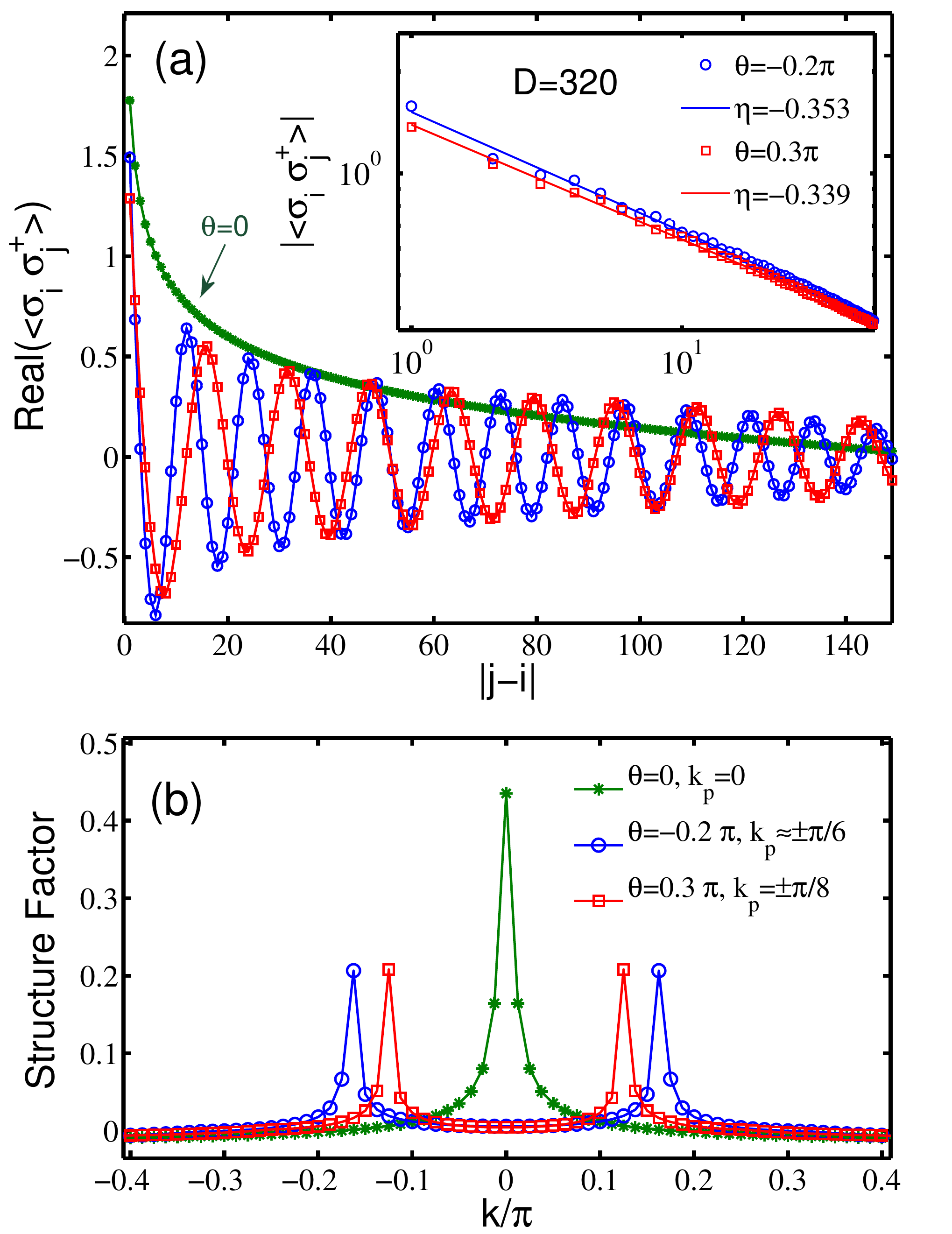}
  \end{center}
  \caption{(a) The spin-spin correlation function (CF) $\langle \sigma_i \sigma^{\dagger}_j \rangle$ exhibits oscillations with varying period depending on $\theta$. The inset (log-log plot) illustrates the algebraic decay of absolute values of CF with distance. (b) The static structure factor of the CFs, revealing explicitly the long period of oscillations.}
  \label{fig:cf_sf}
\end{figure}

The nature of these $c\geq 1$ phases remains unclear. Due to strong finite-size effect we are unable to identify the CFTs of these phases except their central charges. In the following we calculate the spin-spin correlation functions $C(x)\equiv\langle\sigma_i\sigma_{i+x}^\dag\rangle$, which may reveal useful information about the conformal dimensions of the scaling fields in the CFT. We notice that in general the CFT field identification of lattice operators is a highly nontrivial problem~\cite{Mong_Potts}, so caution should be taken in interpreting the numerical results.

Let us start from the $\theta=0,\pi$ exactly solvable points. It has long been known that at $\theta=0$ the $\sigma_i$ operators actually turn into the twist field in the $\mathbb{Z}_3$ parafermion CFT with scaling dimension $\frac{2}{15}$, so $C(x)\sim x^{-4/15}$~~\cite{Mong_Potts}. Similarly, at $\theta=\pi$ the $\sigma_i$ turns into the twist field in the $\mathbb{Z}_4$ parafermion CFT with scaling dimension $\frac{1}{12}$ and $C(x)\sim x^{-1/3}$, which we have verified numerically. In both cases the identification of the continuum limit of $\sigma_i$ is rather straightforward.

Once we move away from the integrable points, the behavior of the spin correlation function becomes more complicated. We find that in the $c=1$ phase, $C(x)$ exhibits oscillations whose characteristic wavevectors depend on $\theta$ [see Fig. \ref{fig:cf_sf}(a)]. This can be seen most easily from the peaks of the static structure factor defined as $S(k) = \sum_{x=1}^{L-1} \cos k x\, C(x)$ [see Fig. \ref{fig:cf_sf}(b)]. This behavior is reminiscent of correlation functions in a Luttinger liquid, which often exhibit oscillations on the scale of Fermi wavelength. We also fit the decay exponent of the envelop function of $C(x)$ for two different values of $\theta$ and in both cases the values are close to $-1/3$. It is tempting to conjecture that the CFT in this phase is closely related to $\mathbb{U}(1)_6$ CFT, but our data is still too preliminary to draw any conclusions. We leave investigations of the CFT for future works.

\section{Conclusions and Discussions}
\label{sec:conclusion}
In this work we numerically study the criticality of a translation-invariant chain of $\mathbb{Z}_N$ parafermion zero modes by mapping to a $\mathbb{Z}_N$ spin model. We completely characterize the low-energy CFT of the $\mathbb{Z}_N$ parafermion chain with NN couplings by a combination of DMRG and ED methods and the results are in perfect agreement with theoretical predictions. We also determine the phase diagram of the $\mathbb{Z}_3$ parafermion chain with up to NNN couplings. We show that the introduction of a relatively small NNN coupling (compared to the NN coupling) can significantly alter the low-energy properties. Phase transitions between different critical phases are also characterized.

We now briefly discuss the physical implications of the results. Parafermion zero modes can be realized at the edge of some Abelian fractional quantum Hall states. For example, by patterning alternating regions gapped out by electron tunneling or s-wave pairing on the edge of a spin-unpolarized $\nu=2/3$ FQH state, $\mathbb{Z}_3$ parafermion zero modes are localized on the domain walls~\cite{Mong_PRX2014} (a similar setup without superconductivity is considered in Ref. [\onlinecite{barkeshli_qubit}]). Virtual tunneling of quasiparticles across the gapped regions then splits the degeneracy, and the effective low-energy Hamiltonian is given by \eqref{eq:ham}. The tunneling amplitudes decay exponentially with the separation between domain walls, i.e. $t_{ij}\sim e^{-|x_i-x_j|/\xi}$, where $x_i$ is the position of the parafermion zero modes and $\xi$ is the correlation length. To the leading approximation, if the domain walls are evenly separated, they collectively realize a $\mathbb{Z}_3$ parafermion CFT. Our results show that $\mathbb{Z}_3$ parafermion CFT is destablized if $t_{\text{NNN}}/t_{\text{NN}}$ is larger than a critical value which we estimate to be $\tan 0.04\pi\approx 0.12$, which roughly corresponds to the separation between NN sites being $\sim 2\xi$.

We also emphasize that the criticality is protected by translation invariance. This should be compared to the topological symmetry  that protects gapless phases in other one-dimensional models of non-Abelian anyonic chains~\cite{GoldenChains, GoldenChain2, Pfeifer_PRB2012}. In fact, one can realize such a $\mathbb{Z}_N$ parafermion chain on the edge of a translation-symmetry enriched topological phase naturally. An exactly solvable model of this type on a square lattice has been recently studied in [\onlinecite{You_PRB2012}]. The topological order in the bulk is identical to that of the $\mathbb{Z}_N$ toric code (or equivalently, a $\mathbb{Z}_N$ lattice gauge theory coupled to matter). However, the translation symmetry has a nontrivial interplay with the topological order: the elementary electric charge and the magnetic charge are exchanged under lattice translations. As a result, $\mathbb{Z}_N$ parafermion zero modes appear on the lattice dislocations. This model also has gapless edge modes if the edge preserves the translation invariance of the system. One can show that the edge can be described by \eqref{eq:ham} exploiting a parafermionic parton representation of the model. The translation symmetry on the edge is inherited from that of the bulk. One might wonder whether the generalized Jordan-Wigner transformation breaks the translation invariance by hand when the parafermion zero modes are grouped to form $\mathbb{Z}_N$ spins~\cite{Teo}. 

\section{Acknowledgment}
M.C. thanks enlightening conversations with Jeffery Teo, David Clarke and Zhenghan Wang and especially Hong-Chen Jiang for very useful discussions on numerical results. H.H.T. thanks Germ{\'a}n Sierra, Michele Burrello and Eddy Ardonne for helpful discussions. M.C. acknowledges the hospitality of the Max-Planck Institute for Quantum Optics in Garching where part of the work was carried out. M.C. acknowledges the hospitality of Perimeter Institute for Theoretical Physics during the finalization of the manuscript. This work has been supported by the DFG through SFB-TR12, the EU project SIQS, and the DFG Cluster of Excellence NIM. Research at Perimeter Institute is supported by the Government of Canada through Industry Canada and by the Province of Ontario through the Ministry of Economic Development and Innovation.

\bibliography{./parafermion}

\appendix
\end{document}